# Regulation of notch sensitivity of lattice materials by strut topology

K. Li[a], P. E. Seiler[a], V. S. Deshpande[a], N. A. Fleck[a,*]

[a] *Department of Engineering, University of Cambridge, Cambridge CB2 1PZ, United Kingdom*

**Abstract**

We propose a local reinforcement technique for lattices in the vicinity of a stress-raiser such as a notch, in order to elevate the macroscopic strength and ductility. A spatially non-uniform waviness distribution of sinusoidally-shaped struts is assumed in the vicinity of the notch, and the sensitivity of macroscopic tensile response to strut waviness distribution is studied by finite element analysis. Optimized lattice structures are determined in order to maximise the macroscopic tensile strength or ductility from these various strut waviness distributions. Both hexagonal and triangular lattices are studied as these geometries are representative of bending-dominated and stretching-dominated lattices, respectively.

## 1. Introduction

Periodic networks of struts, called lattice materials, are increasingly used in a large variety of engineering applications, e. g. tower structures in civil engineering, the cores of lightweight sandwich panels, microscopic mechanical filters [1], and soft network materials utilized in bio-integrated electronics [2]. Recent studies have mostly focused on the design, fabrication and modelling of perfect lattice materials. However, in some practical applications, lattice materials contain local stress-raisers such as holes [3], notches [4, 5] and solid inclusions [6]. Ronan et al. [6] explored the sensitivity of strength and ductility of a regular honeycomb to the notch, and found that the crack-like irregular cell, as generated by missing cell walls, can significantly knock-down the macroscopic ductility and strength of the hexagonal lattice. In order to reinforce a lattice

*Corresponding author

*Email addresses:* `kl513@cam.ac.uk` (K. Li), `pes34@cam.ac.uk` (P. E. Seiler), `vsd@eng.cam.ac.uk` (V. S. Deshpande), `naf1@eng.cam.ac.uk` (N. A. Fleck)

material in the presence of a stress-raiser, we propose the use of a spatially non-uniform waviness distribution of struts in the vicinity of the notch.

### *1.1. Classes of lattice*

Lattices can be classified as bending-dominated structures, such as the hexagonal lattice, or stretching-dominated structures, such as the triangular lattice [7]. On a lower scale of structural hierarchy, each strut can exhibit a bending or stretching response depending upon its shape; for example, a sinusoidal shape confers high axial compliance due to induced bending under an axial load. These hierarchical lattice structures can be sub-divided as follows [8]:

(i) *stretching lattice:* A triangular lattice with straight struts is stretching-dominated, both on the lattice scale or on the strut scale. This lattice has a high macroscopic modulus and inherits the ductility of the cell wall material [9].

(ii) *stretching-bending lattice:* The macroscopic stiffness of a stretching-dominated lattice (such as the triangular lattice) is reduced by increased axial compliance of the struts due to waviness.

(iii) *bending lattice:* A hexagonal lattice with straight or wavy struts deforms by bending of the struts.

### *1.2. Influence of strut waviness on macroscopic properties*

Wavy struts have a larger axial ductility than straight struts due to the additional macroscopic strain associated with straightening of the struts. Consequently, a lattice made from wavy struts may possess a significantly enhanced ductility. In support of this assertion, Ma et al. [2, 10] and Jang et al. [11] found that polyimide lattice materials, comprising horseshoe-shaped struts embedded in a soft polymeric matrix, possess a significantly enhanced ductility. Moreover, the macroscopic modulus [11] and Poisson's ratio [12] can be modulated by the introduction of wavy struts. For example, Symons and Fleck [13] and Grenestedt [14] have predicted the reduction in macroscopic stiffness of triangular lattices due to waviness of the struts.

### 1.3. Influence of imperfections on lattices

As-fabricated lattices can possess manufacturing defects such undercuts [15] and Plateau borders at the nodes [6], a random positioning of the nodes, missing cell-walls, and a variation in strut thickness. These defects affect the macroscopic properties of the lattices to a varying degree, as discussed by Simone and Gibson [16], Chen et al. [17], Zhu et al. [18, 19], Fleck and Qiu [20], Romijn and Fleck [21], Ronan et al. [6], and Seiler et al. [15].

Additionally, sub-components made from lattice materials may contain geometric features, such as holes, notches and solid inclusions that act as stress-raisers. For example, Liu et al. [3] investigated the ductility of a soft network geometry containing a hole to accommodate hard, inorganic electronic components. The presence of such stress-raisers can significantly knock-down the overall ultimate tensile strength and ductility for both stretching-dominated and bending-dominated lattices.

### 1.4. Scope of study

The purpose of this study is to control the macroscopic tensile response of lattice materials by the choice of strut topology in the vicinity of a stress-raiser such as a notch. Each strut has a sinusoidal shape but the amplitude can vary strut by strut (spatially graded wavy). Part A reports a combined experimental and finite element (FE) study on the tensile response of low carbon steel hexagonal and triangular lattices, manufactured by rapid prototyping, with wavy struts present near a central notch. The sensitivity of the macroscopic tensile response to the presence of waviness is demonstrated. In Part B, a parametric FE study is performed on a representative volume element (RVE) of a periodic lattice (hexagonal and triangular) containing a notch and the influence of waviness distribution of the struts upon the macroscopic tensile response is investigated. A design map is constructed to show the effect of waviness and its spatial extent upon macroscopic strength and ductility. Optimal designs are identified in order to maximise the ultimate tensile strength or ductility.

## 2. Part A: An assessment of reinforcement near a notch by local waviness (experiments and FE)

### 2.1. Tests on notched lattices

The potency of graded wavy lattices to ameliorate the effects of a stress raiser in a lattice upon strength and ductility is revealed by preliminary experiments. Specimens were water-jet cut from hot-rolled steel sheets of grade *S275* (low carbon steel with a maximum of 0.25% C by weight) of Vickers hardness 170HV30. The sheets were of thickness $B = 3\,\text{mm}$. Hexagonal and triangular lattice specimens with straight struts (Figs. 1(a) and (b)) or wavy struts (Figs. 2(a) and (b)) were water-jet cut with a centre-notch of length $2a$ resulting from $n_\text{b}$ missing struts. The zone of wavy struts was limited to the vicinity of the centre-notch, with a range of values of maximum waviness. The relative density of the as-manufactured hexagonal and triangular lattices was $\bar{\rho} = 0.17 \pm 0.003$, where

$$\bar{\rho} = \alpha \frac{t}{\ell}, \tag{1}$$

in terms of strut thickness $t$, strut length $\ell$ while the coefficient $\alpha = 2\sqrt{3}$ and $2/\sqrt{3}$ for the triangular and hexagonal lattices [22], respectively

Sketches of lattices (gauge area of width $W$ and length $H$) with straight struts are presented in Figs. 1(a) and (b), and those with graded waviness close to the centre notch are shown in Figs. 2(a) and (b). The gauge area $W \times H$ was of size $156\,\text{mm} \times 140\,\text{mm}$ for the hexagonal lattices (comprising $17 \times 10$ cells) and was of size $212\,\text{mm} \times 184\,\text{mm}$ for the triangular lattices (comprising $16 \times 12$ cells). These lattices were generated using the code given in [23].

With $\zeta$ measured along the straight strut from one end Fig. 2(a), the initial deflection $\delta$ of the strut (in unloaded state), from its straight configuration, is given by

$$\delta = A \sin\left(\frac{2\pi\zeta}{\ell}\right) \tag{2}$$

where $A$ is the amplitude of the wavy strut and $\ell$ the distance between the end points of the strut, i. e. the length of the strut in the limit $A \to 0$. The distribution of amplitude of waviness $A$ of struts in the vicinity of the centre notch is given by

$$\frac{A}{\ell} = \max\left(0, \frac{A_\mathrm{m}}{\ell}\left[1 - 5.4\left(\frac{x_\mathrm{c}}{W}\right)^2 - 16\left(\frac{y_\mathrm{c}}{H}\right)^2\right]\right) \quad (3)$$

where $A_\mathrm{m}$ is the maximum amplitude, and $x_\mathrm{c}$ and $y_\mathrm{c}$ are the Cartesian coordinates of the centre of struts measured with respect to the centre of the specimen. We emphasise that this distribution of the waviness amplitude has been chosen based on the ability to manufacture specimens: here we report the ability of the FE simulations to capture experimental measurements and then in Section 3 present a detailed FE parametric study where we determine the optimal amplitude distribution. Both triangular and hexagonal lattices comprise the same strut thickness $t = 0.75\,\mathrm{mm}$ but different strut lengths ($\ell_\mathrm{hex} = 5.0\,\mathrm{mm}$ and $\ell_\mathrm{tria} = 15.3\,\mathrm{mm}$) such that the relative density of the lattice absent the notch is $\bar{\rho} = 0.17$. The increased waviness near the notch leads to greater strut arc lengths between the joints, and consequently the effective relative densities of the triangular and hexagonal lattices are slightly increased (see Seiler et al. [15] for details).

The as-manufactured hexagonal and triangular lattices also contained defects in the form of undercuts, of depth $e$, and Plateau borders of radius $r_\mathrm{n}$; see Fig. 1. The influence of these defects on the macroscopic behaviour has been reported recently by Seiler et al. [15]. In the present study, X-ray computer tomography (CT)[1] of the mid-plane of the manufactured lattices revealed that the Plateau border radius has a normal distribution with a mean value $\bar{r}_\mathrm{n} = 0.11\,\mathrm{mm}$ and a standard deviation $\Delta r_\mathrm{n} = 0.01\,\mathrm{mm}$; this is expressed via the notation $r_\mathrm{n} = 0.11 \pm 0.01\,\mathrm{mm}$. Likewise, the undercut depth was measured as $e = 0.10 \pm 0.04\,\mathrm{mm}$, and the strut thickness of manufactured lattices was $t = 0.80 \pm 0.33\,\mathrm{mm}$. The reported mean values of the strut thickness, undercut depth, and Plateau border radius were used to create the FE models.

[1] *X-TEK, XT H 225ST*, voxel size: 30 µm.

### 2.2. Material characterization

Macroscale dogbone specimens of the parent material were manufactured to characterise the solid material properties. The material properties of the solid, low-carbon steel sheets were measured from a large dogbone-shaped specimen tested at nominal tensile strain rates of $2 \times 10^{-4}\,\mathrm{s}^{-1}$ and $2 \times 10^{-3}\,\mathrm{s}^{-1}$. Preliminary tests were done with the loading direction either aligned with the rolling direction or transverse to the rolling direction; the degree of anisotropy was negligible (within a few percent), and so the tensile response is reported only for tests with the loading direction aligned with the rolling direction. The true stress versus true strain response of solid dogbone specimens is only mildly sensitive to the strain rate in the above range (see Fig. 3) and therefore rate effects are neglected in the current study. The Young's modulus of the solid material, as measured from the dogbone specimen, is $E_\mathrm{s} = 210 \pm 12\,\mathrm{GPa}$. The 0.2% offset yield strength is $\sigma_\mathrm{YS} = 338 \pm 12\,\mathrm{MPa}$, the ultimate tensile strength is $\sigma_\mathrm{UTS} = 500 \pm 6\,\mathrm{MPa}$, and nominal tensile failure strain is $\varepsilon_\mathrm{fs} = 0.24 \pm 0.01$. The measured true stress versus true strain response is shown up to maximum load in Fig. 3, and the extrapolated response was used in the FE analysis.

### 2.3. Tensile tests on notched lattices

The tensile response of the lattices were conducted using a screw-driven test machine, and Digital Image Correlation (DIC) to measure displacements. All lattices were tested in uniaxial tension. The macroscopic nominal strength and ductility for lattices are defined in terms of the measured load $P$ and the extension $U$ of the gauge section of length $H$ (Fig. 1). Results are presented in terms of a nominal stress $P/(WB)$ and nominal strain $U/H$. All lattices were tested at a nominal strain rate of $\dot{\varepsilon} = \dot{U}/H = 2 \times 10^{-4}\,\mathrm{s}^{-1}$. The nominal stress versus strain response of notched specimens with graded waviness is shown in Fig. 4(a) for hexagonal lattices and in Fig. 4(b) for triangular lattices with the deformed lattices shown in Figs. 4(c) and (d). The solid black lines represent the measured tensile response of lattices comprising straight struts ($A_\mathrm{m}/\ell = 0$) without a centre notch, as well as those with a length of the centre notch of 4 cells (resulting from $n_\mathrm{b} = 4$ missing struts in hexagonal lattices and $n_\mathrm{b} = 3$ in triangular lattices, see Figs. 1(a) and (b)) and peak amplitude $A_\mathrm{m}/\ell$ from 0 (straight struts) to 0.075.

Consider first the response of the hexagonal lattices. The graded waviness increases the macroscopic ductility of notched samples along with a minor change in tensile strength. Consequently, the (nominal) energy absorption capacity, as estimated by the area under the $P/(WB)$ versus $U/H$ curve, is enhanced by the presence of strut waviness. In contrast, additional waviness in a triangular lattice has a negligible effect upon macroscopic ductility, but knocks down the strength. Thus, the addition of waviness degrades the energy absorption capacity of the triangular lattice. The tensile strength $P_{\mathrm{f}}$ of the lattices normalized by the strength of the un-notched sample $P_0$ is shown in Fig. 5(a) as a function of $A_{\mathrm{m}}/\ell$. Likewise, the macroscopic ductility $U_{\mathrm{f}}/H$ is plotted as a function of $A_{\mathrm{m}}/\ell$ in Fig. 5(b). These plots emphasize the trends already noted in Fig. 4: the addition of graded waviness near the notch in a hexagonal lattice leads to a large increase in $U_{\mathrm{f}}/H$ and to a smaller increase in $P_{\mathrm{f}}/P_0$. In contrast, for the triangular lattice, $U_{\mathrm{f}}/H$ is almost insensitive while $P_{\mathrm{f}}/P_0$ drops significantly with increasing $A_{\mathrm{m}}/\ell$.

### *2.4. Finite element simulations*

Finite element simulations were performed using ABAQUS/Standard v2018 to simulate the tensile response of the lattices under uniaxial tension. All struts had identical defects with a Plateau border radius and undercut depth equal to the mean value measured from the midsection of CT images of as-manufactured specimens (see Section 2.1). Uniaxial loading of the finite lattice was simulated by constraining all degrees of freedom along the bottom edge of the specimen while the top edge was subjected to uniform displacement in the $y$-direction of the specimen. The FE mesh of the lattice comprised 8-noded, plane-strain elements with quadratic shape functions (type CPE8). A mesh sensitivity study revealed that adequate accuracy is achieved by placing at least 4 elements across the thickness of each strut.

J2 flow theory was assumed, with the tensile stress $\sigma$ versus strain $\varepsilon$ response of the cell wall solid given by the measured true stress versus true strain relation shown in Fig. 3. The finite element simulations were taken to the point where one of the struts of the lattice begins to neck. A neck was defined as a reduction of the strut thickness by 20% compared to its original thickness. Excellent agreement is noted between the FE predictions and the measured responses, see Figs. 4 and 5. This gives confidence in the use of FE simulation to model the response of the lattices containing both as-manufactured defects and macroscopic notches.

The increased strength and ductility of hexagonal, bending-dominated lattices by the presence of local waviness suggests that the addition of graded waviness near stress-raisers can give structural benefit. In the following, the potency of graded waviness is explored by a more general FE-based parametric study of the distribution of strut amplitude, for a range of notch lengths in both hexagonal and triangular lattices.

## 3. Part B: The search for an optimal distribution of waviness

### *3.1. Finite Element Analysis*

A comprehensive set of finite element simulations were performed using ABAQUS/Standard v2018 to simulate the tensile response of the lattices under uniaxial tension. The 2D geometries for FE models are defined from the Pythonscript interface of Rhinoceros v 6[2]. Plateau borders radius $r_\mathrm{n} = 0.11$ are included for all struts, but undercuts are not included in the simulations.

### *3.2. Parametric study and optimisation*

A periodic Representative Volume Element (RVE) of lattice is studied, for both hexagonal and triangular lattices, see Figs. 6 and 7. The non-uniform distribution of strut waviness of RVE is illustrated in Fig. 6(b) and Fig. 7(b). The wavy struts are sinusoidal in shape, with an amplitude $A$ defined by the Gaussian spatial distribution

$$A(r) = A_\infty + (A_\mathrm{m} - A_\infty) \cdot \exp\left(-\frac{r^2}{\lambda^2}\right) \tag{4}$$

where $r$ is the distance from the centroid of a strut to the centre of RVE, such that $r^2 = x^2 + y^2$. This Gaussian distribution makes use of 3 non-dimensional parameters $A_\infty/\ell, A_\mathrm{m}/\ell$ and $\lambda/\ell$. The dependence of the macroscopic strength and ductility on these three parameters was mapped via an extensive FE investigation with the optimal locations for performing FE calculations in the design space chosen via the *Optimal Latin Hypercube sampling* method [24]. The design space spanned $A_\infty/\ell$ and $A_\mathrm{m}/\ell \in [0, 0.2]$ and $\lambda/\ell \in [0, 20]$ for the hexagonal lattice. The corresponding space for the triangular lattice was $A_\infty/\ell$ and $A_\mathrm{m}/\ell \in [0, 0.06]$ and $\lambda/\ell \in [0, 20]$. Design maps with contours were constructed from the discrete FE results using a Gaussian fitting procedure. Further, a refined optimisation was performed near the minima/ maxima in strength and ductility estimated using the Gaussian fitting to accurately define optimum parameters. This local optimisation employed a NLPQL (Nonlinear Programming by Quadratic Lagrangian) algorithm [25].

[2] https://www.rhino3d.com

### *3.3. Tensile response of lattices with graded waviness*

The preliminary experimental and numerical assessments in Part A of the current study have demonstrated that graded waviness around a notch modulates the macroscopic tensile responses of the lattices. In the experimental assessment, the strut thickness was held fixed at $t = 0.75\,\text{mm}$. Consequently, the relative density of the lattice increases locally by the presence of strut waviness. In this parametric FE study (Part B), we held the relative density fixed at $\bar{\rho} = 0.17$ by varying the strut thickness $t$ of each strut such that $t\ell_{\text{s}}$ = constant for each strut where $\ell_{\text{s}}$ is the arc length of the strut. Uniaxial tensile loading of the periodic RVE was simulated by imposing an axial strain in the $y$-direction. The FE mesh and material used in the simulations is the same as described in Section 2.4.

#### *3.3.1. Hexagonal lattices*

The square RVE (Fig. 6(a)) of side length $2D$ ($D/\ell = 12$) contains a central notch of length $2a$, such that $a/\ell = 3.5$ corresponds to a notch with 4 missing struts ($n_{\text{b}} = 4$) and $a/\ell = 0.9$ corresponds to $n_{\text{b}} = 1$. The waviness of struts varies throughout the RVE in accordance with Eq. (4), where $r$ is the distance from the centroid of the strut to the centre of RVE (Fig. 6(b)). Based on a set of 80 such simulations, contours of the normalized macroscopic tensile strength $\hat{\sigma}$ and ductility $\varepsilon_{\text{f}}^{\infty}$ of the hexagonal lattice are plotted in Fig. 8 as a function of the 3 non-dimensional groups $\lambda/\ell$, $A_{\infty}/\ell$, and $A_{\text{m}}/\ell$ for the cases of $n_{\text{b}} = 1$ and 4.
Here, strength $\hat{\sigma}$ is defined as

$$\hat{\sigma} = \frac{\sigma_{\text{UTS}}(A_{\infty}/\ell, A_{\text{m}}/\ell, \lambda/\ell, n_{\text{b}})}{\sigma_{\text{UTS}}\,(0, 0, \lambda/\ell, n_{\text{b}})} \tag{5}$$

where the nominator $\sigma_{\text{UTS}}(A_{\infty}/\ell, A_{\text{m}}/\ell, \lambda/\ell, n_{\text{b}})$ is the peak nominal stress for any choice of imperfection parameters; the denominator is the corresponding strength of the lattice with the same notch but straight struts. The ductility $\varepsilon_{\text{f}}^{\infty}$ is defined by the nominal strain at which $\sigma_{\text{UTS}}$ is attained. According to the design maps of Figs. 8(a) and (b), the strength of the notched hexagonal lattice is maximised for a finite zone of waviness near the notch and straight struts ($A_\infty/\ell = 0$) remote from the notch. Alternatively, the strength is minimised by employing straight struts near the notch $A_\mathrm{m}/\ell = 0$ but wavy remote struts with $A_\infty/\ell > 0$. This quantitative behaviour persists for hexagonal lattices with $n_\mathrm{b} = 1$ and 4, but the optimal values of $(A_\infty/\ell, A_\mathrm{m}/\ell$ and $\lambda/\ell)$ is dependent upon the value of $n_\mathrm{b}$. The ductility design maps for hexagonal lattices are given in Figs. 8(c) and (d); they reveal that, for both $n_\mathrm{b} = 1$ and $n_\mathrm{b} = 4$, the macroscopic ductility increases with the strut waviness either near the notch or remotely. The geometric parameters that optimise strength and ductility in the design region are summarized in Table 1.

It is clear from both Figs. 8(a) and (b) that the notch strength is maximised by the choice $A_\infty/\ell = 0$, and by intermediate, finite values of $A_\mathrm{m}/\ell$ and $\lambda/\ell$. The key results are summarized in Fig. 9(a) by plotting $\hat{\sigma}$ versus $A_\mathrm{m}/\ell$, with $A_\infty/\ell = 0$ and $\lambda/\ell = 1.5$ for $n_\mathrm{b} = 1$, and $\lambda/\ell = 4$ for $n_\mathrm{b} = 4$. These choices of $A_\infty/\ell$ and $\lambda/\ell$ define lines in $(A_\infty/\ell, A_\mathrm{m}/\ell,\ \lambda/\ell)$ space that pass through the maximum value of $\hat{\sigma}$. Note that the hexagonal lattice with a larger notch requires a greater waviness $A_\mathrm{m}/\ell$, and a larger wavy region radius $\lambda/\ell$ to obtain maximum strength. The corresponding trajectory of macroscopic ductility $\varepsilon_\mathrm{f}^\infty$ versus $A_\mathrm{m}/\ell$ is shown in Fig. 9(b), indicating that the ductility increases with maximum waviness $A_\mathrm{m}/\ell$ in a monotonic manner.

It is instructive to determine the sensitivity of tensile response to the degree of waviness by considering selected cases. The nominal stress versus strain responses of the periodic hexagonal lattices is shown in Fig. 10, for the choice of waviness that maximises and minimises strength in addition to the reference case of straight struts, for the notch size ($n_\mathrm{b} = 4$). We note that the presence of graded waviness can significantly modulate the macroscopic tensile response of hexagonal lattices. The design that maximises the strength also slightly increases the ductility, thereby increasing the energy absorbing capability. The design that minimises the strength $\hat{\sigma}$ significantly increases the ductility.

The achievable design space for the hexagonal lattice, in terms of maximum and minimum strength $\hat{\sigma}$ and ductility $\varepsilon_\mathrm{f}^\infty$, is plotted in Fig. 11 for graded waviness within the regime $A_\infty/\ell, A_\mathrm{m}/\ell \in [0, 0.2]$ and $\lambda/\ell \in [0, 10]$. By designing a hexagonal lattice with graded waviness in this design region, the macroscopic strength can be increased up to 6% for $n_\mathrm{b} = 1$ and 26% for $n_\mathrm{b} = 4$. On the other side, the macroscopic strength can also be as low as 35% and 52% that of the straight strut lattices, for $n_\mathrm{b} = 1$ and 4, respectively (Fig. 11(a)). The macroscopic ductility of the hexagonal lattice can also be significantly modulated (Fig. 11(b)). For example, for $n_\mathrm{b} = 1$, the ductility ranges from 0.29 to 0.82 in relation to $\varepsilon_\mathrm{f}^\infty = 0.32$ for the reference design with straight struts. Similarly for $n_\mathrm{b} = 4$, $\varepsilon_\mathrm{f}^\infty$ ranges from 0.23 to 0.78, in comparison to 0.27 for the straight-strut design.

Insight into the effect of waviness upon the failure mode is gleaned from the undeformed and deformed configurations of hexagonal lattices as shown in Fig. 12, for $n_\mathrm{b} = 4$. Three designs are shown: the straight strut design $A_\infty/\ell = A_\mathrm{m}/\ell = 0$ (Fig. 12(a)), the maximised strength design $A_\infty/\ell = 0, A_\mathrm{m}/\ell = 0.15, \lambda/\ell = 4$ (Fig. 12(b)) and the minimised strength design $A_\infty/\ell = 0.2, A_\mathrm{m}/\ell = 0, \lambda/\ell = 4$ (Fig. 12(c)). The deformed configurations are shown at peak load for that geometry. Contours of normalized von Mises effective stress $\sigma_\mathrm{e}$ are included in the form of $\bar{\sigma} = \sigma_\mathrm{e}/\sigma_\mathrm{YS}$. The stress distribution directly ahead of the notch is sensitive to the waviness distribution with the key findings being:

(i) For the hexagonal lattice with straight struts, the strut at the notch tip fails first (Fig. 12(a)), followed by the adjacent strut, with increasing applied strain.

(ii) For the graded waviness design that maximises the strength, the set of vertical struts directly ahead of the notch root fail almost simultaneously, as is evident by the high stresses $\bar{\sigma}$ and the onset of necking as shown in Fig. 12(b).

(iii) The design that minimises the strength has straight struts around the notch, and remote waviness (Fig. 12(c)). Here, the struts adjacent to the notch fail first, while the wavy struts next to them are not stretched out completely and are far from reaching the tensile necking strain.

### *3.3.2. Triangular lattices*

A similar study has been performed on triangular lattices, with a representative square periodic RVE of size $2D \times 2D$ ($D/\ell = 12$) and notches of length $a/\ell = 0.6$ and 1.7, corresponding to $n_\mathrm{b} = 1$ and 4, respectively (Fig. 7). The graded waviness of struts is again defined by Eq. (4). The design maps of Fig. 13 show normalized macroscopic tensile strength and ductility for triangular lattices over a wide range of waviness distribution $A_\infty/\ell, A_\mathrm{m}/\ell \in [0, 0.06], \lambda/\ell \in [0, 10]$, for the cases of a notch as characterized by $n_\mathrm{b} = 1$ and $n_\mathrm{b} = 4$.

The imposition of graded waviness onto the triangular lattice degrades the tensile strength for $n_\mathrm{b} = 1$ and only leads to a 1% increase in tensile strength for $n_\mathrm{b} = 4$, see Figs. 13(a) and (b). This is in contrast to the significant elevation in tensile strength for the hexagonal lattice, recall Figs. 8 and 9. As for the hexagonal lattices, the ductility design maps shown in Figs. 13(c) and (d) suggest that the overall ductility of triangular lattices increases with increasing $A_\infty$ or $A_\mathrm{m}$. The geometric parameters that maximise strength and ductility of the triangular lattice are summarized in Table 2. Further, the sensitivity of strength and ductility to the amplitude of waviness $A_\mathrm{m}/\ell$ is shown in Fig. 14 for $n_\mathrm{b} = 1$ and $n_\mathrm{b} = 4$, with $A_\infty = 0$ and $\lambda/\ell = 1.5$. The values of $A_\infty/\ell$ and $\lambda/\ell$ are chosen such that this trajectory of fixed $A_\infty/\ell$ and $\lambda/\ell$ runs through the optimal design associated with maximum tensile strength of the triangular lattice.

The macroscopic nominal stress versus strain responses of triangular lattices with straight struts, and with graded waviness that maximises or minimises the strength over the design range, are included in Fig. 15 for the choice $n_\mathrm{b} = 4$. The graded waviness design that maximises $\hat{\sigma}$ exhibits a stress versus strain curve that is almost identical to that of the lattice with straight struts. In contrast, the stress-strain curve for the minimum strength design with $n_\mathrm{b} = 4$ has a much reduced initial yield strength and only a slightly enhanced ductility.

The range in strength and ductility of triangular lattices in the feasible design region $(A_\infty/\ell, A_\mathrm{m}/\ell) \in [0, 0.06], \lambda/\ell \in [0, 10]$ is shown in Fig. 16. The normalized tensile strength $\hat{\sigma}$ ranges from 0.92 to 1.0 for $n_\mathrm{b} = 1$, and from 0.93 to 1.01 for $n_\mathrm{b} = 4$ (see Fig. 16(a)); the range of achievable ductility for triangular lattice is 0.115 to 0.147 for $n_\mathrm{b} = 1$, and 0.105 to 0.142 for $n_\mathrm{b} = 4$. We conclude that the presence of graded waviness is detrimental to both strength and ductility except for a limited choice of waviness for which the improvement in strength or ductility is minor. The deterioration in strength and ductility is associated with an increased stress concentration at the notch root when graded waviness is present; this was deduced from deformed meshes (of similar type to that shown in Fig. 12 for the hexagonal mesh) that are omitted here for the sake of brevity.

## 4. Concluding remarks

The present study has revealed, by a combination of experiment and finite element analysis, that the presence of graded waviness near a notch can have a major effect upon the collapse response. By a suitable choice of the amplitude and size of wavy domain, in relation to the notch length, it is possible to increase both the strength and ductility significantly for a hexagonal lattice but not for a triangular lattice. A similar approach can be adopted for other types of stress-raiser such as a hole or inclusion. Other local reinforcement strategies are possible: whilst the addition of local waviness may not change the local relative density of the lattice, it is also possible to adopt other strategies such as a change in strut thickness that will change the distribution of relative density. More substantial changes to lattice configuration are also possible: lattice size and connectivity can be modified near a stress-raiser, analogous to the refinement of a finite element mesh near a stress-raiser. These more sophisticated approaches as well as investigations of the control of notch sensitivity for other notch geometries and lattice topologies such as auxetic lattices await future studies.

## Acknowledgement

The authors gratefully acknowledge the financial support from the European Research Council (ERC) under the European Union's Horizon 2020 research and innovation program, grant GA669764, MULTILAT.

**Tables**

| | $n_\mathrm{b}=1$ $\hat{\sigma}$(max) | $n_\mathrm{b}=1$ $\hat{\sigma}$(min) | $n_\mathrm{b}=4$ $\hat{\sigma}$(max) | $n_\mathrm{b}=4$ $\hat{\sigma}$(min) | $n_\mathrm{b}=1$ $\varepsilon_\mathrm{f}^\infty$(max) | $n_\mathrm{b}=1$ $\varepsilon_\mathrm{f}^\infty$(min) | $n_\mathrm{b}=4$ $\varepsilon_\mathrm{f}^\infty$(max) | $n_\mathrm{b}=4$ $\varepsilon_\mathrm{f}^\infty$(min) |
|---|---|---|---|---|---|---|---|---|
| $A_\infty/\ell$ | 0 | 0.2 | 0 | 0.2 | 0 | 0.2 | 0 | 0.2 |
| $A_\mathrm{m}/\ell$ | 0.1 | 0 | 0.15 | 0 | 0 | 0.2 | 0 | 0.2 |
| $\lambda/\ell$ | 1.5 | 1.5 | 4.0 | 4.0 | NA | NA | NA | NA |

**Table 1:** Geometric parameters for optimized graded waviness distributions of hexagonal lattices in the design region $A_\infty/\ell, A_\mathrm{m}/\ell \in [0, 0.2]$ and $\lambda/\ell \in [0, 10]$, with notch size $n_\mathrm{b} = 1$ and 4.

| | $n_\mathrm{b}=1$ $\hat{\sigma}$(max) | $n_\mathrm{b}=1$ $\hat{\sigma}$(min) | $n_\mathrm{b}=4$ $\hat{\sigma}$(max) | $n_\mathrm{b}=4$ $\hat{\sigma}$(min) | $n_\mathrm{b}=1$ $\varepsilon_\mathrm{f}^\infty$(max) | $n_\mathrm{b}=1$ $\varepsilon_\mathrm{f}^\infty$(min) | $n_\mathrm{b}=4$ $\varepsilon_\mathrm{f}^\infty$(max) | $n_\mathrm{b}=4$ $\varepsilon_\mathrm{f}^\infty$(min) |
|---|---|---|---|---|---|---|---|---|
| $A_\infty/\ell$ | 0 | 0.06 | 0 | 0.06 | 0 | 0.06 | 0 | 0.06 |
| $A_\mathrm{m}/\ell$ | 0 | 0 | 0.04 | 0 | 0 | 0.06 | 0 | 0.06 |
| $\lambda/\ell$ | NA | 1.5 | 1.5 | 1.5 | NA | NA | NA | NA |

**Table 2:** Geometric parameters for optimized graded waviness distributions of triangular lattices in the design region $A_\infty/\ell, A_\mathrm{m}/\ell \in [0, 0.06]$ and $\lambda/\ell \in [0, 10]$, with notch size $n_\mathrm{b} = 1$ and 4.

**Figures**

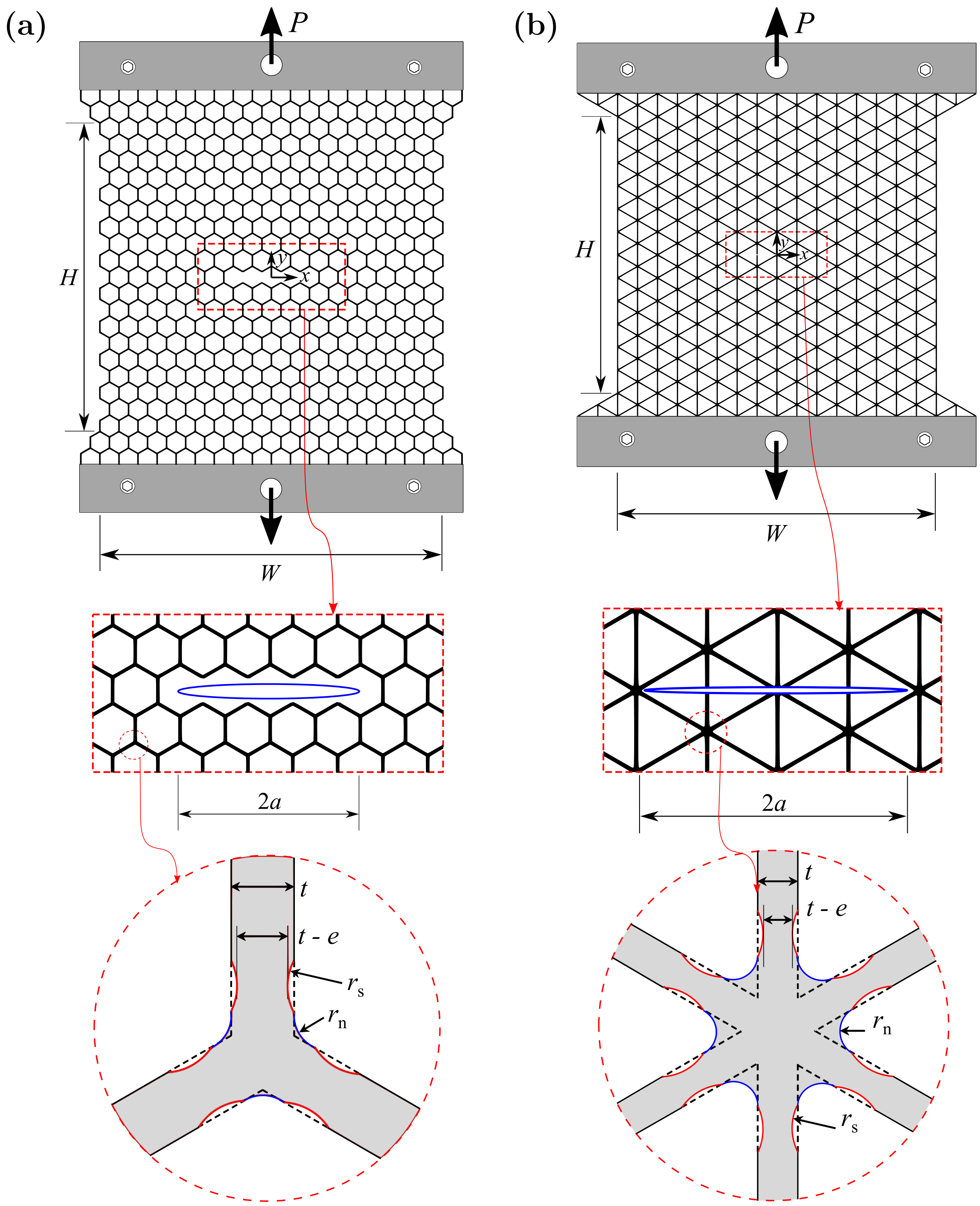

**Figure 1:** Sketch of experimental lattice specimens ($\bar{\rho} = 0.17$) containing as-designed defects in the form of a row of missing cell walls for (a) hexagonal and (b) triangular lattices.

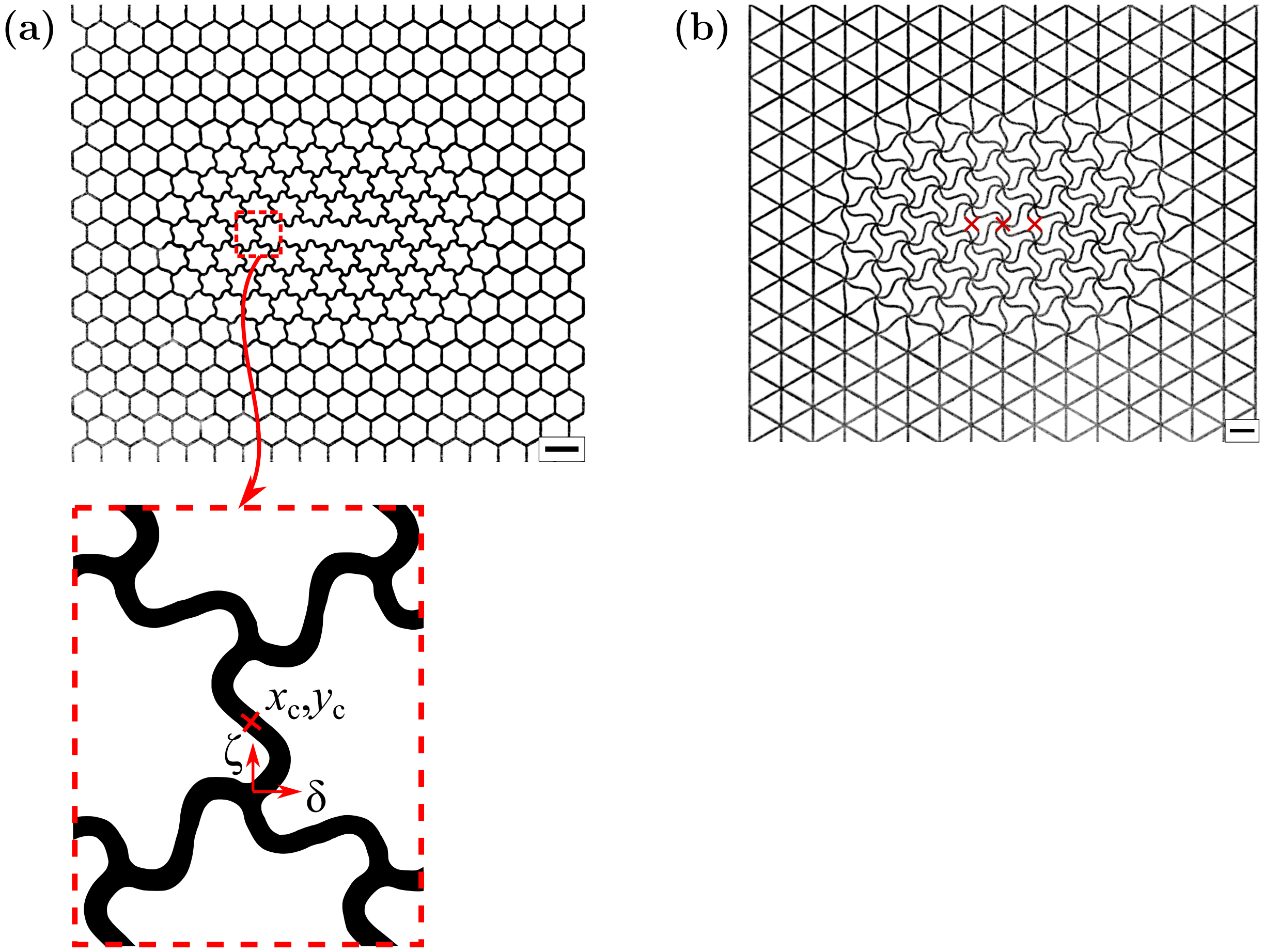

**Figure 2:** Specimens ($\bar{\rho} = 0.17$) with graded wavy struts around a centre notch of (a) hexagonal lattices and (b) triangular lattices. In (b), the red crosses denote cut struts prior to testing. The scale bars are of length 10 mm.

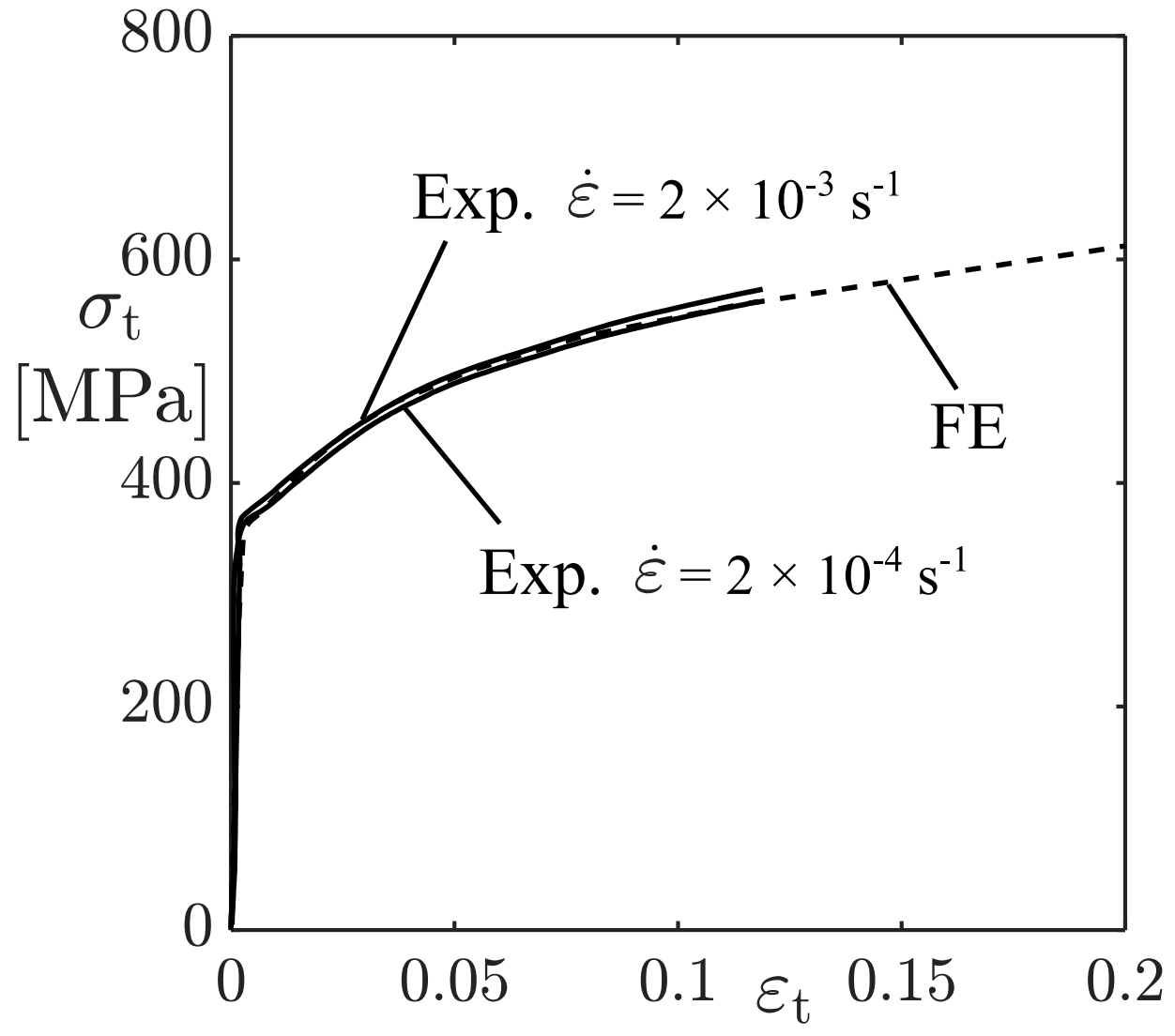

**Figure 3:** True stress versus true strain response of solid dogbone shaped specimens and the assumed response used in the finite element (FE) analysis which includes the used extrapolation.

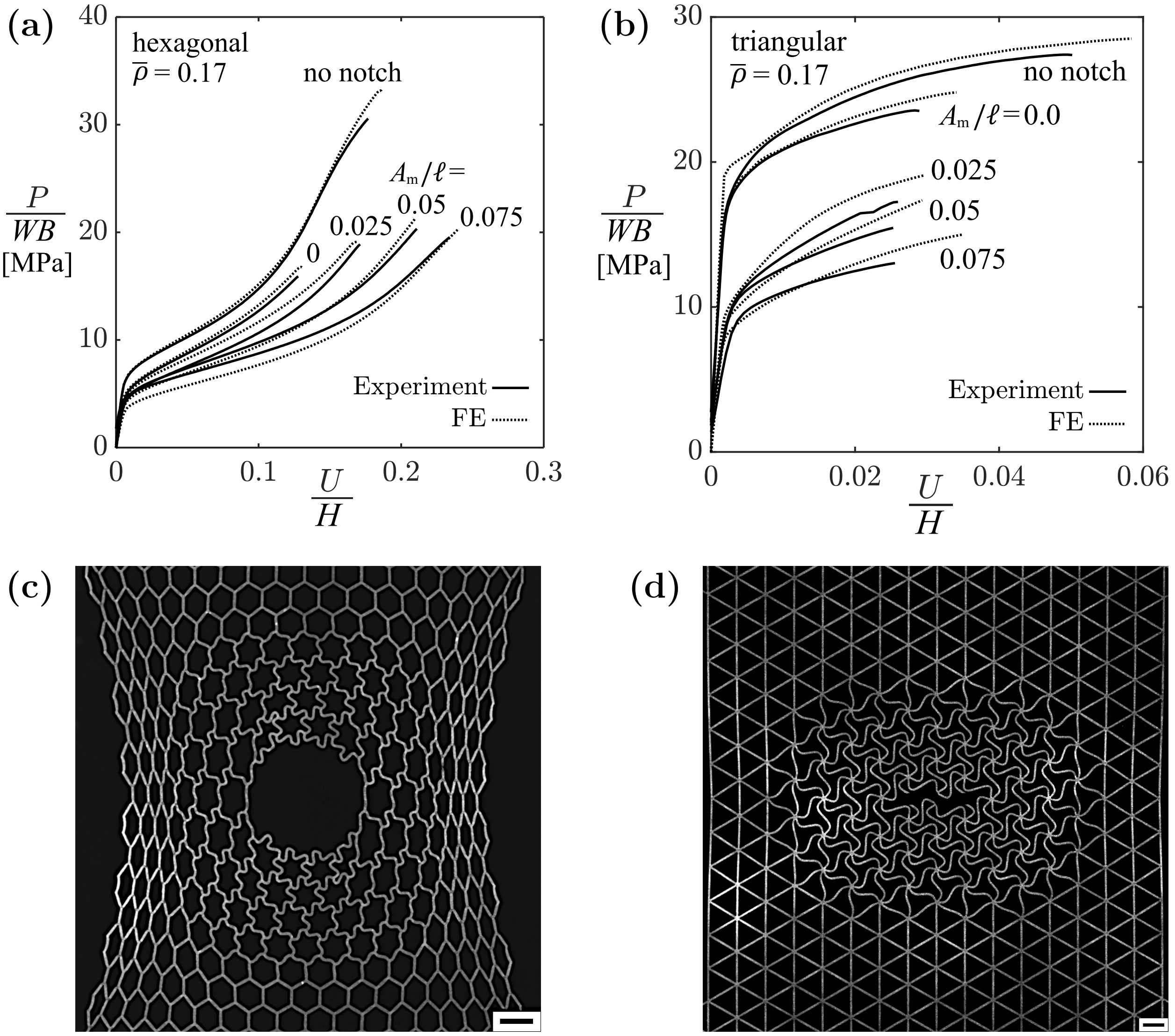

**Figure 4:** Measured stress versus strain response of (a) hexagonal and (b) triangular lattices. Photographs of the deformed (c) hexagonal and (d) triangular lattices ($\bar{\rho} = 0.17$, $A_m/\ell = 0.075$) before first strut failure. The scale bars are of length 10 mm.

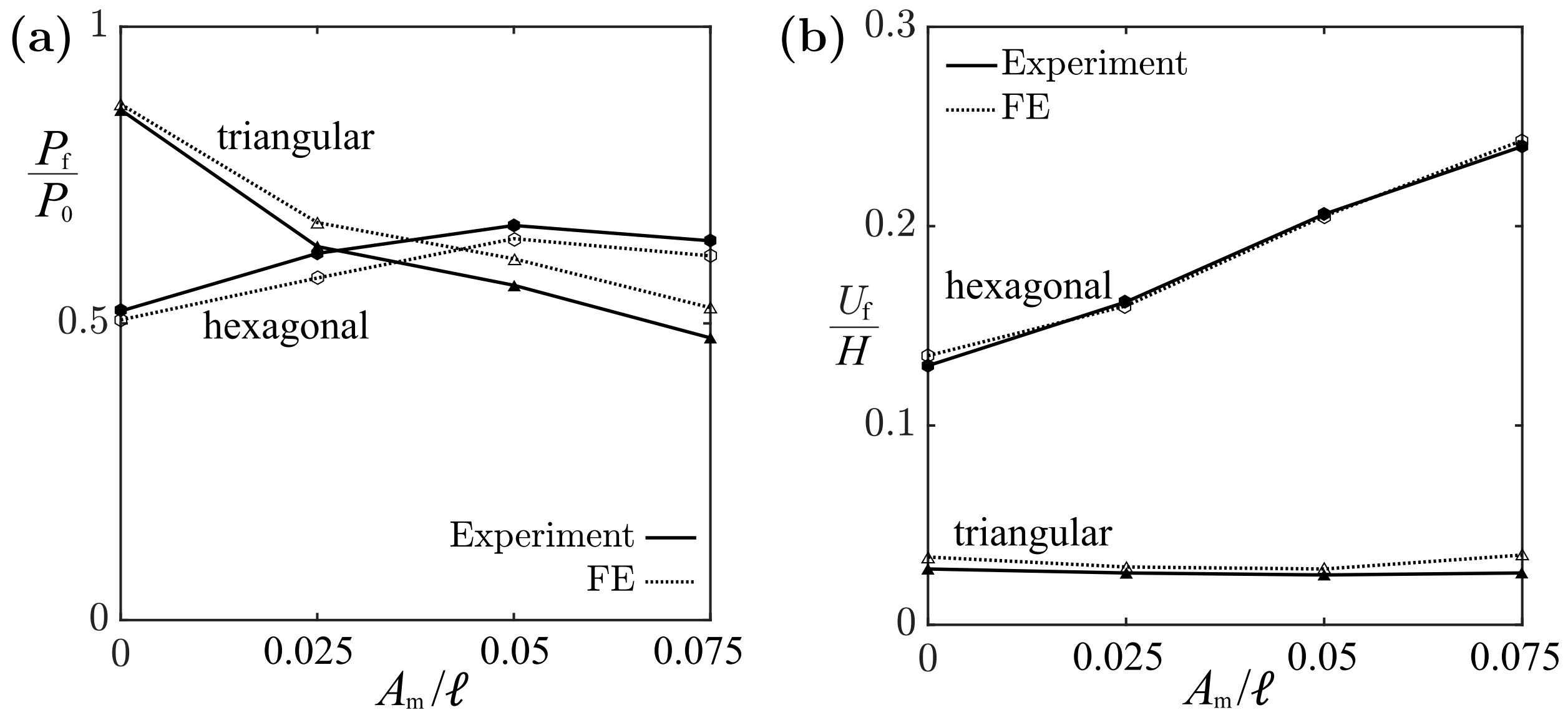

**Figure 5:** (a) Normalised gross strength $P_f/P_0$ and (b) ductility $U_f/H$ at first strut failure versus the maximal amplitude $A_m$ of graded wavy hexagonal and triangular lattices with a length of the centre notch of 4 cells.

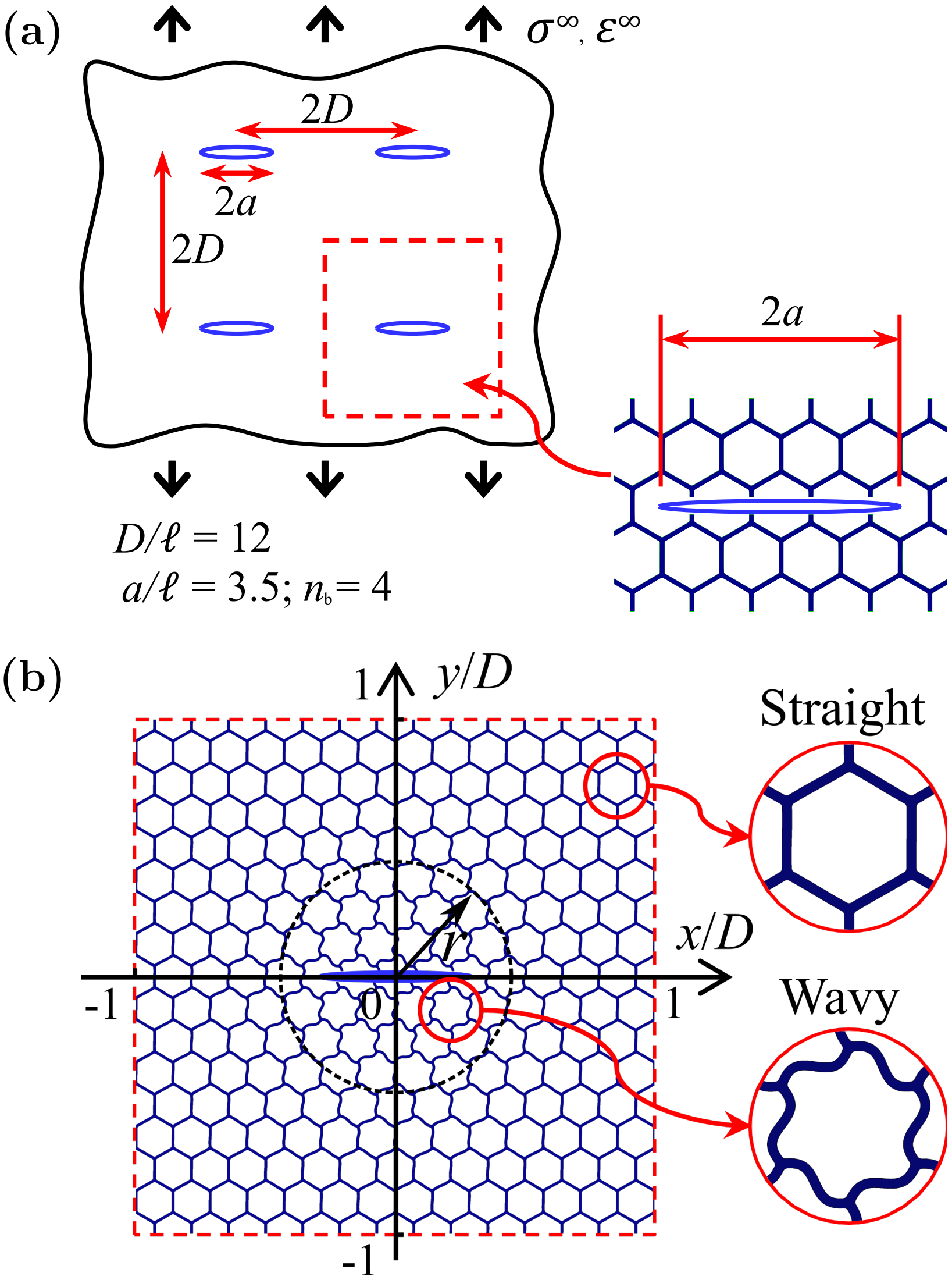

**Figure 6:** Hexagonal lattices with a periodic distribution of flaws. (a) A flaw of length $2a$, consisting of missing cell walls, exists in a periodic RVE of size $2D \times 2D$. (b) Geometry of a periodic RVE of hexagonal lattice with graded strut waviness.

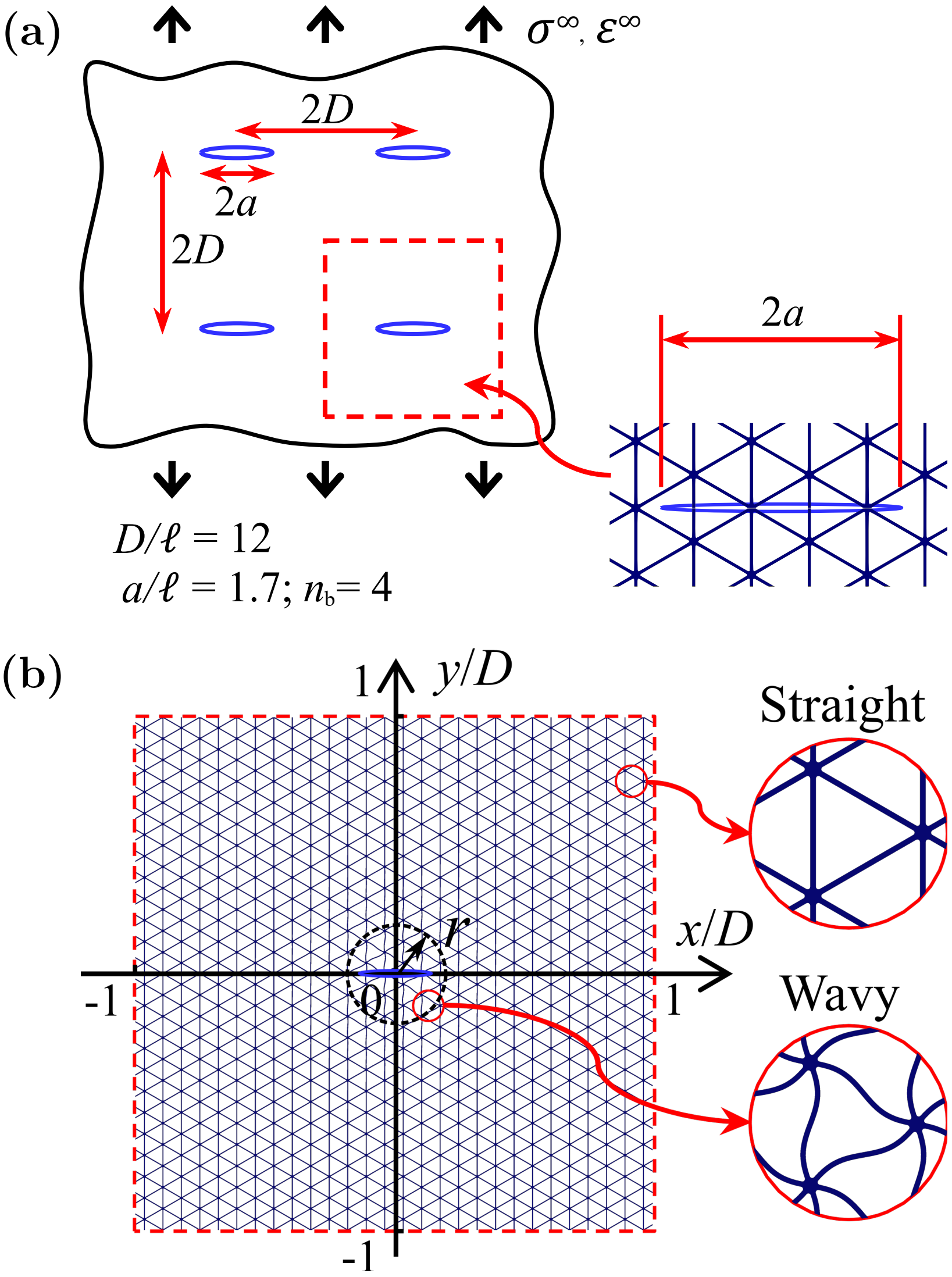

**Figure 7:** Triangular lattices with a periodic distribution of flaws. (a) A flaw of length $2a$, consisting of missing cell walls, exists in a periodic RVE of size $2D \times 2D$. (b) Geometry of a periodic RVE of triangular lattice with graded strut waviness.

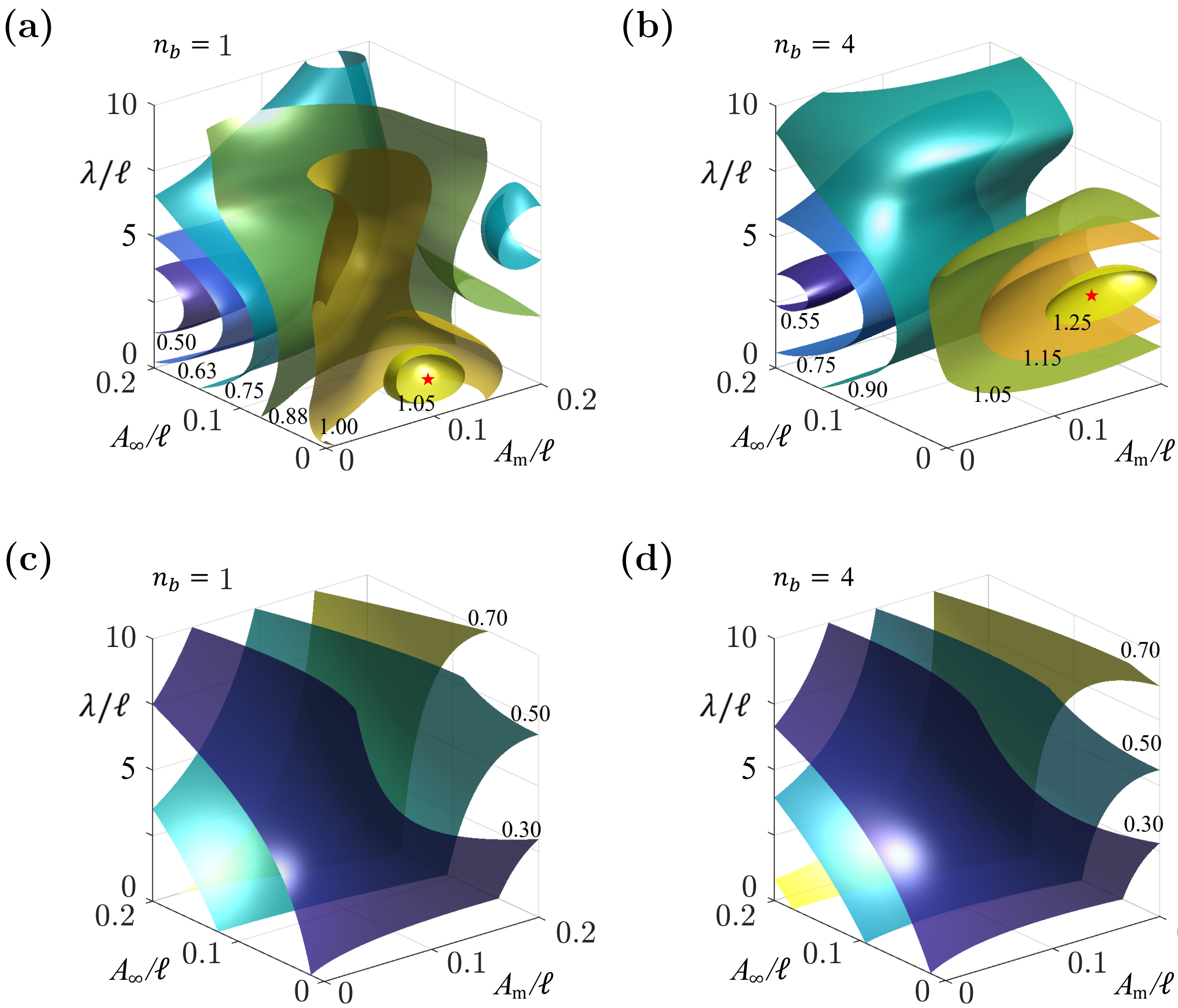

**Figure 8:** Influence of graded strut waviness on ultimate tensile strength and ductility of hexagonal lattices. Effect of $A_\infty/\ell$, $A_m/\ell$ and $\lambda/\ell$ on normalised UTS $\hat{\sigma}$ for hexagonal lattices with notch sizes (a) $n_\mathrm{b} = 1$ and (b) $n_\mathrm{b} = 4$, and on ductilities $\varepsilon_\mathrm{f}^\infty$ of lattices with (c) $n_\mathrm{b} = 1$ and (d) $n_\mathrm{b} = 4$.

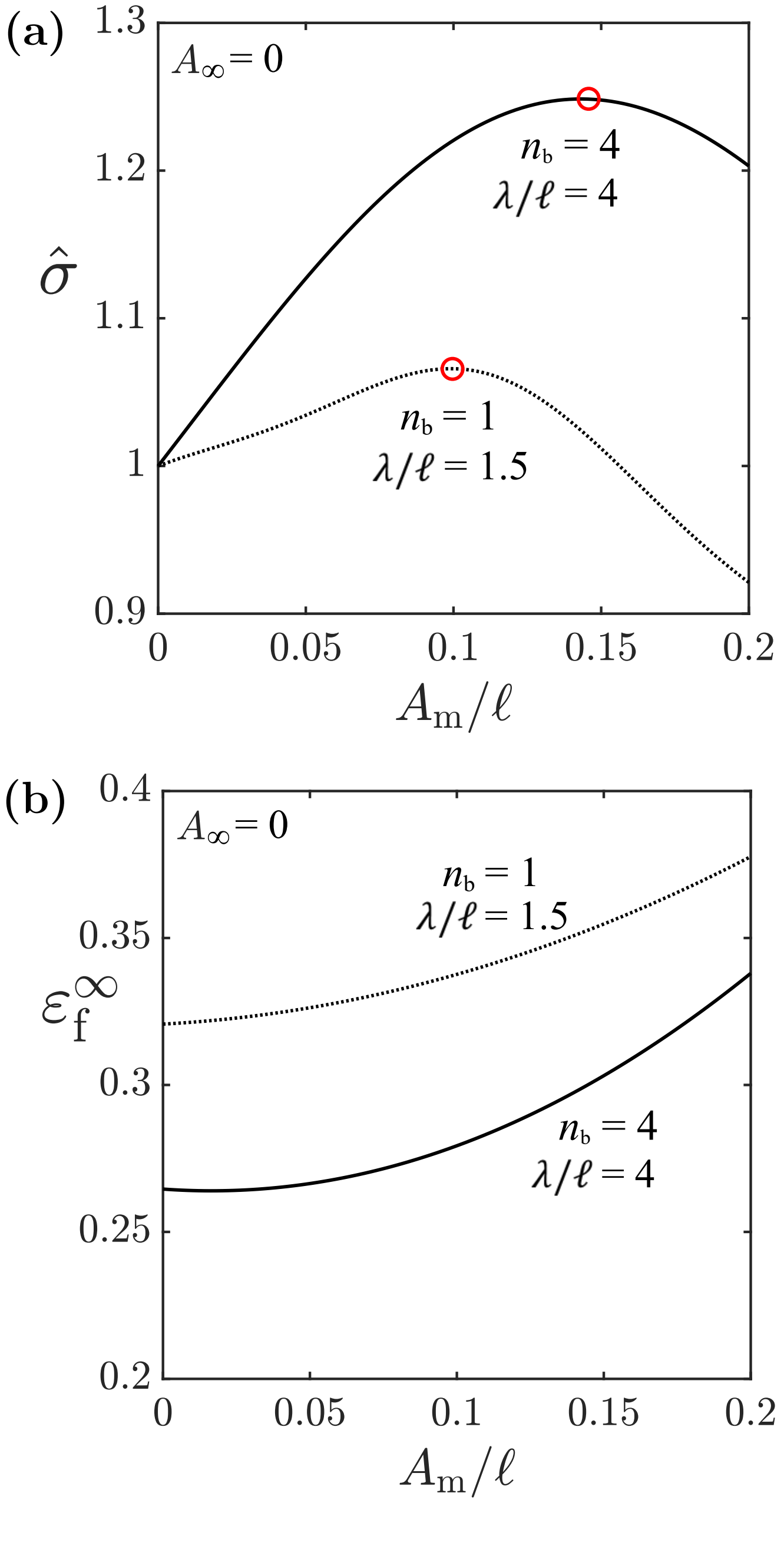

**Figure 9:** Effect of waviness amplitude $A_\mathrm{m}$ upon (a) strength $\hat{\sigma}$ and (b) ductility $\varepsilon_\mathrm{f}^\infty$ of a hexagonal lattice for the case of straight remote struts ($A_\infty = 0$).

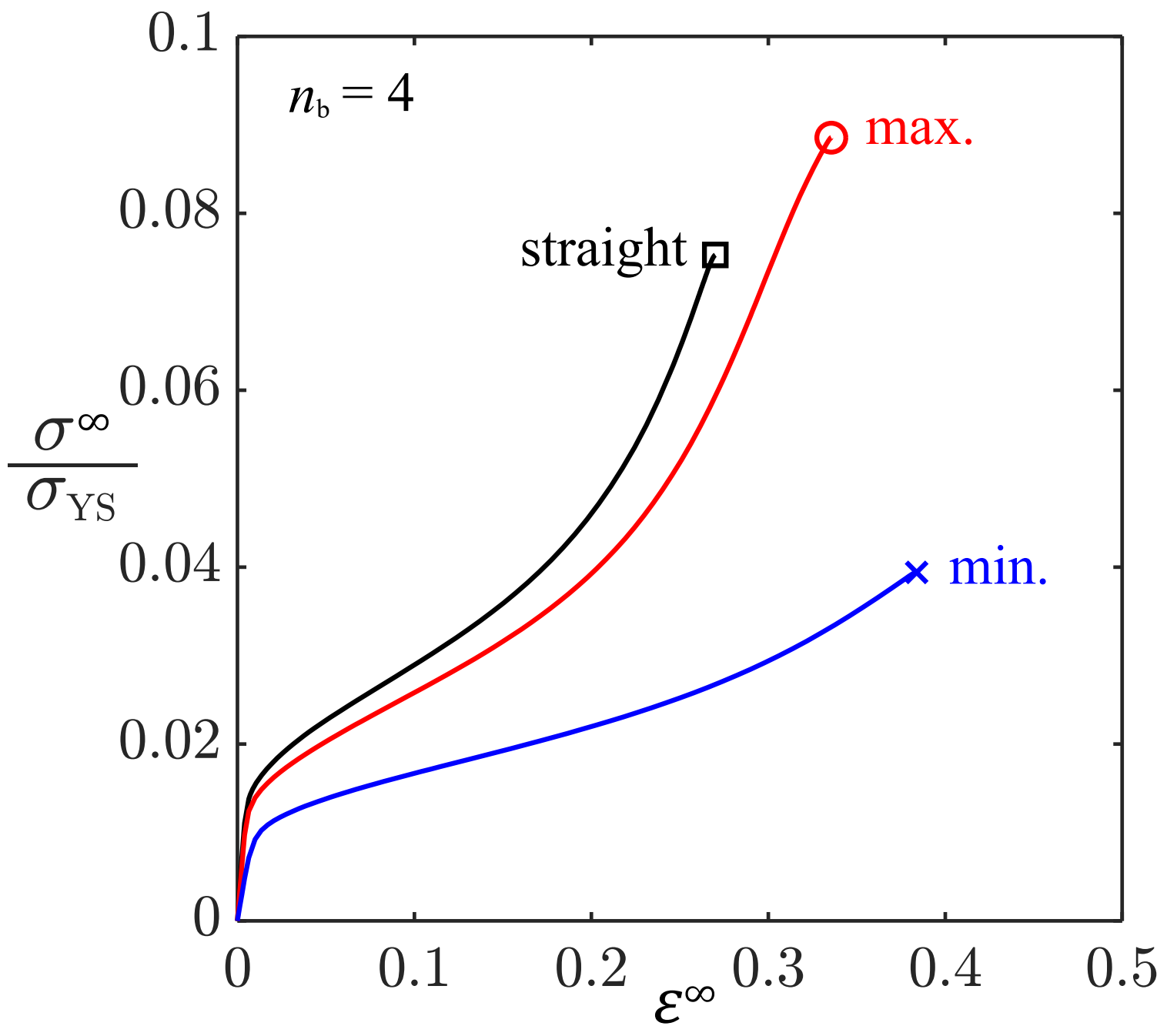

**Figure 10:** Tensile responses for periodic hexagonal lattices with missing cell walls ($n_b = 4$) of designed graded strut waviness distributions; the lattice is modulated to have the maximum and minimum UTS in the design region $A_\infty/\ell, A_m/\ell \in [0, 0.2]$ and $\lambda/\ell \in [0, 10]$.

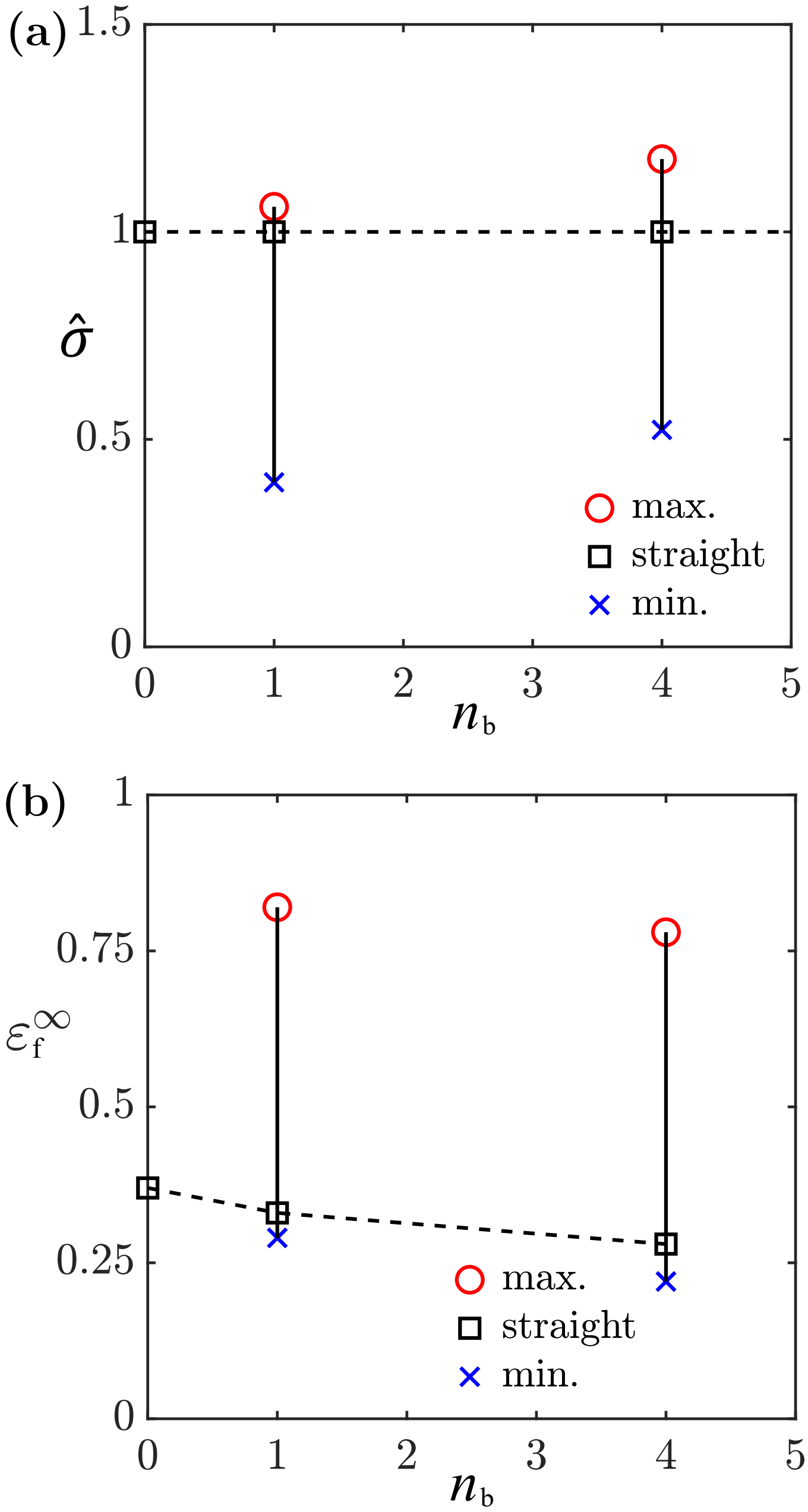

**Figure 11:** The achievable range of (a) strength and (b) ductility of periodic hexagonal lattices, for a strut waviness in the feasible design region $A_\infty/\ell, A_m/\ell \in [0, 0.2]$ and $\lambda/\ell \in [0, 10]$.

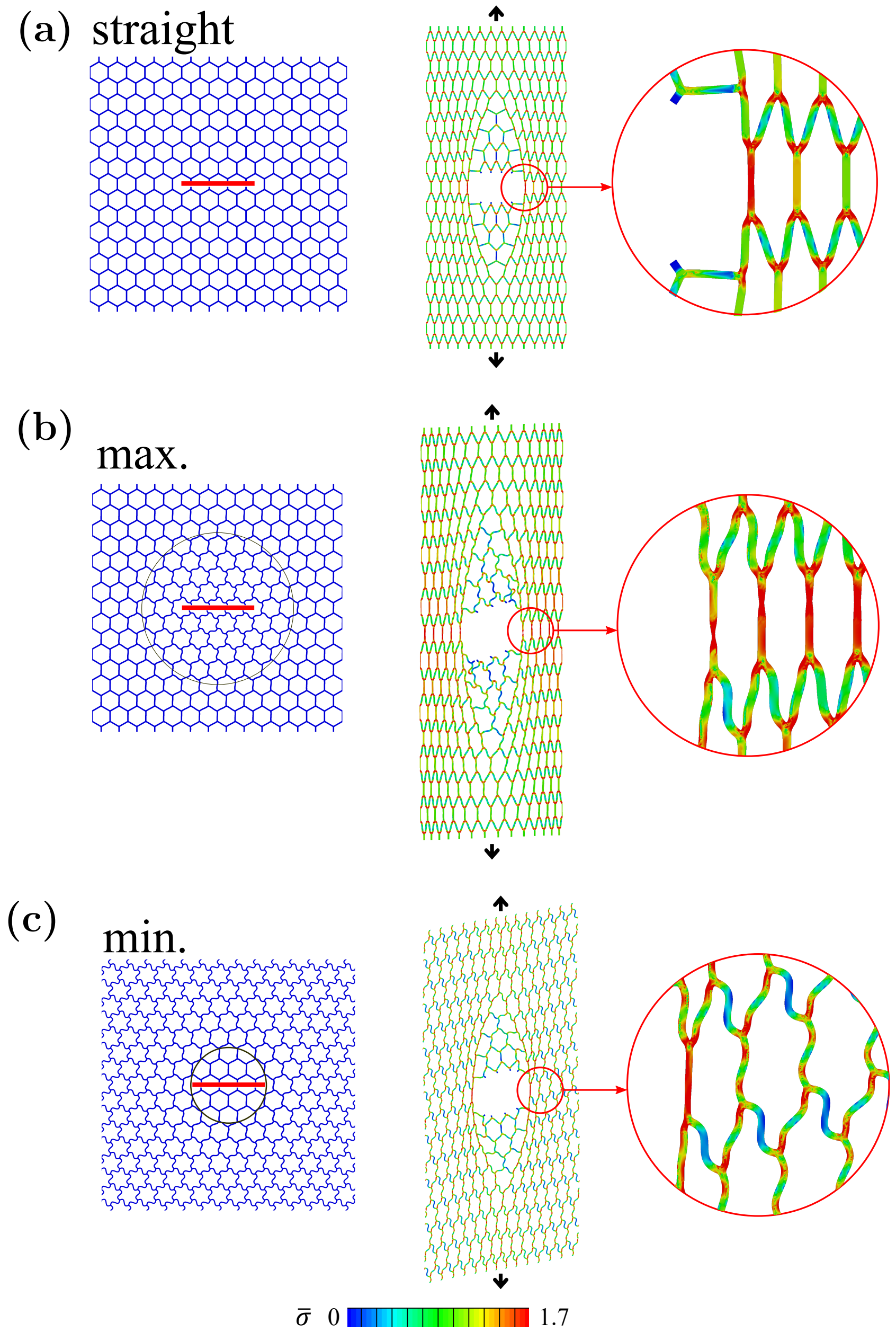

**Figure 12:** Undeformed and deformed configuration of periodic hexagonal lattices of $n_\mathrm{b} = 4$ with (a) original straight struts, and graded waviness distribution with (b) maximum and (c) minimum strengths. The black circle is the boundary for wavy and straight struts, where $\frac{A-A_\infty}{A_m-A_\infty} = 0.1$. The contour value in the defined meshes is the normalised von Mises stress with respect to the yield stress ($\bar{\sigma} = \sigma/\sigma_{\mathrm{YS}}$).

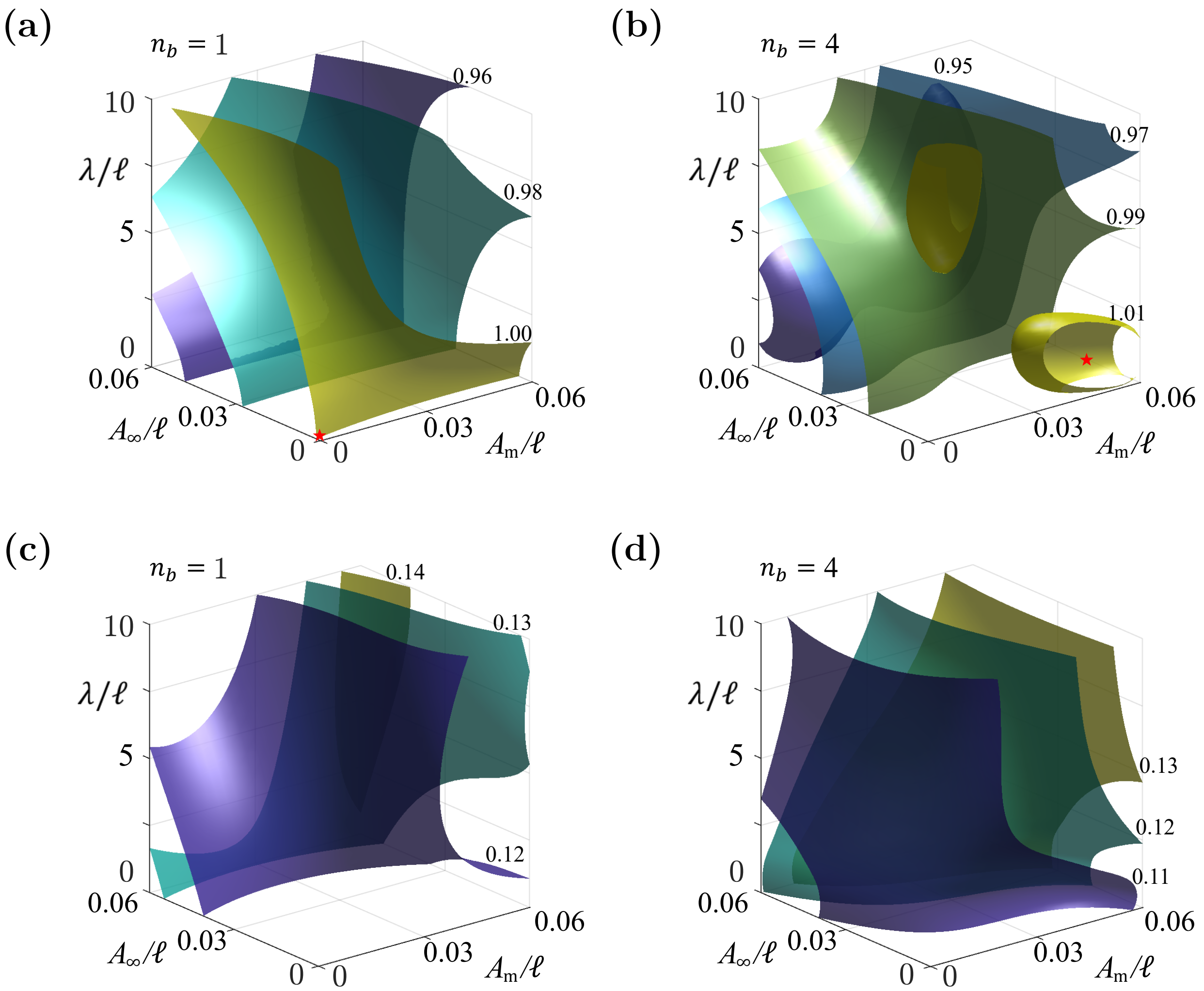

**Figure 13:** Influence of strut waviness distribution on ultimate tensile strength and ductility of triangular lattices. Effect of $A_\infty/\ell$, $A_m/\ell$ and $\lambda/\ell$ on normalised UTS $\hat{\sigma}$ for triangular lattices with notch sizes (a) $n_b = 1$ and (b) $n_b = 4$, and on ductilities $\varepsilon_f^\infty$ of lattices with (c) $n_b = 1$ and (d) $n_b = 4$.

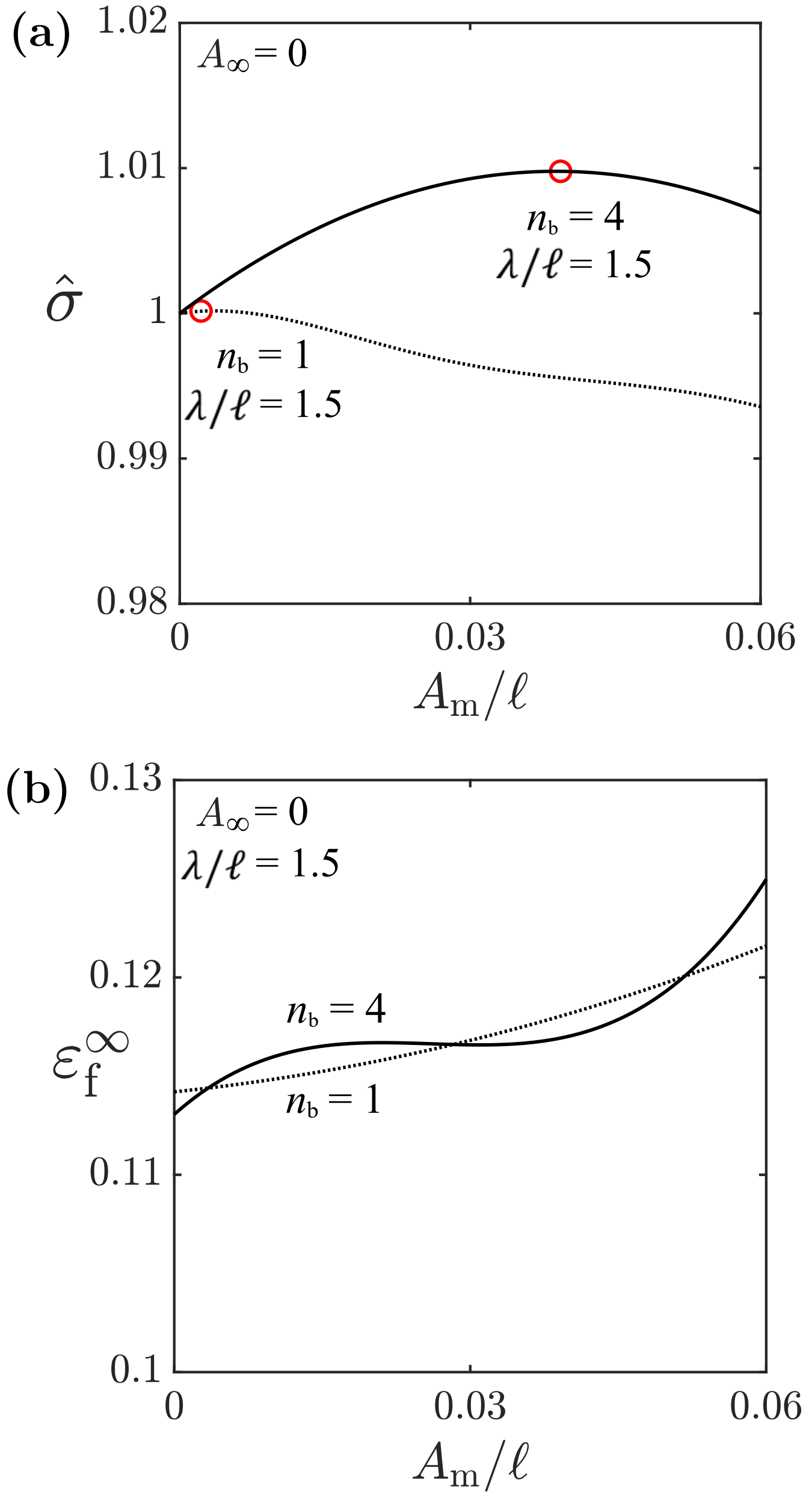

**Figure 14:** Effect of waviness amplitude $A_\mathrm{m}$ upon (a) strength $\hat{\sigma}$ and (b) ductility $\varepsilon_\mathrm{f}^\infty$ of a triangular lattice for the case of straight remote struts ($A_\infty = 0$).

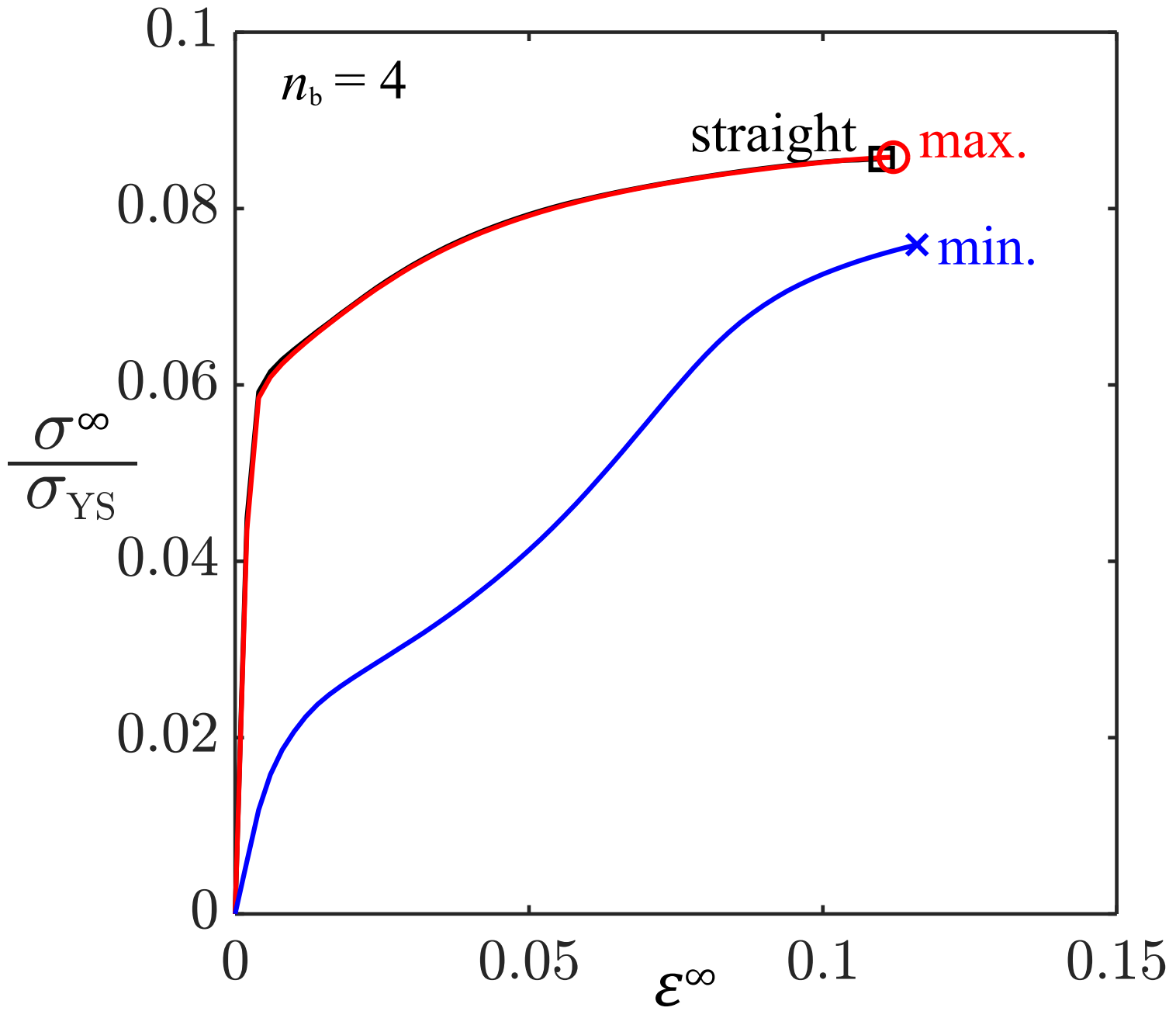

**Figure 15:** Tensile responses of periodic triangular lattices with missing cell walls ($n_b = 4$) of designed graded strut waviness distributions. The lattice has the maximum and minimum UTS in the design region $A_\infty/\ell, A_m/\ell \in [0, 0.06]$ and $\lambda/\ell \in [0, 10]$, with the case if straight struts $A_\infty = A_m = 0$ included for reference.

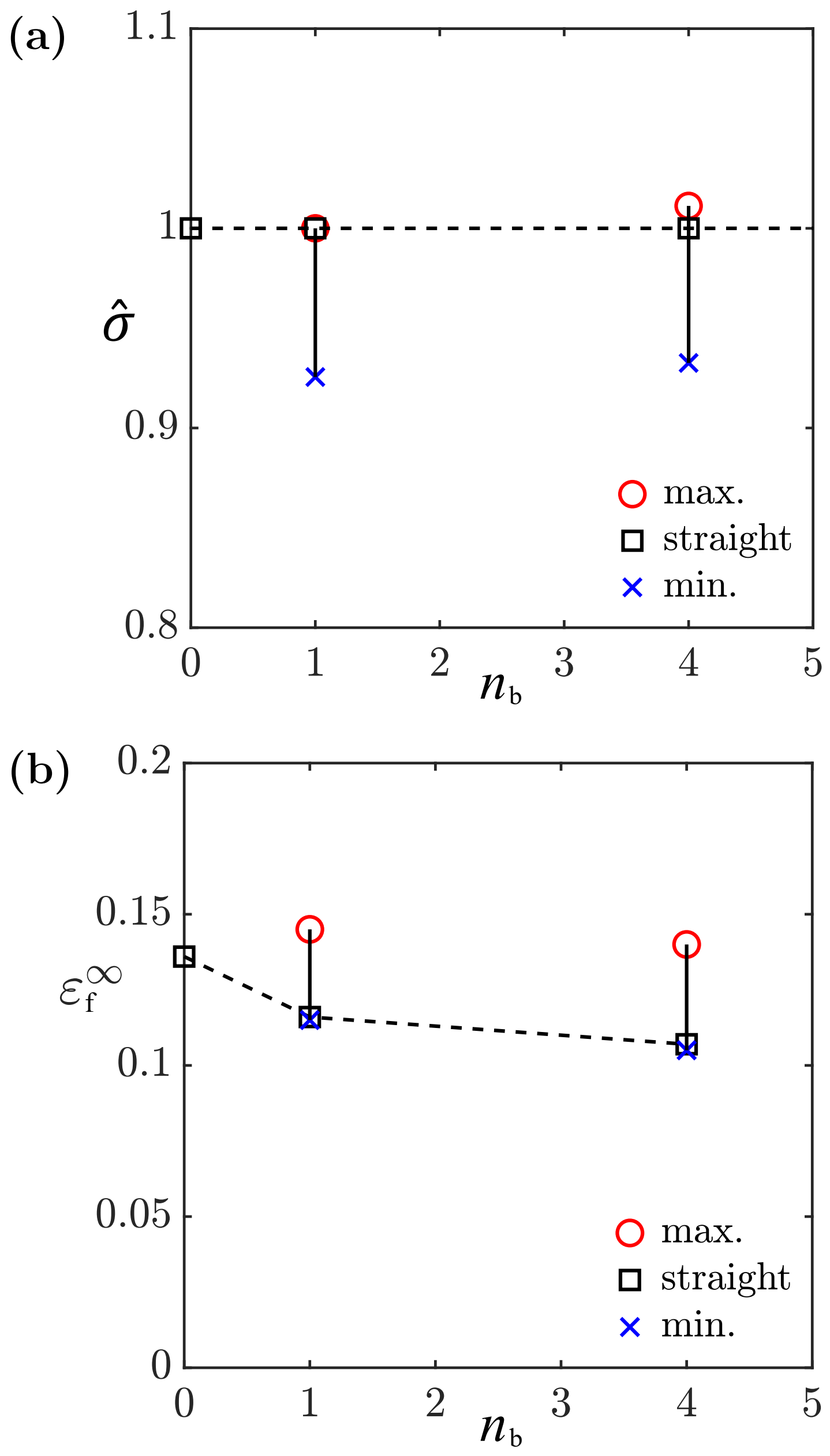

**Figure 16:** The achievable range of (a) strength and (b) ductility of periodic triangular lattices, for a strut waviness in the feasible design region $A_\infty/\ell, A_m/\ell \in [0, 0.06]$ and $\lambda/\ell \in [0, 10]$.